\documentclass[twoside]{article}
\usepackage{amssymb,amsmath}
\usepackage{qic}
\usepackage{bbm}
\usepackage{comment}

\begin{document}
\setlength{\textheight}{8.0truein}    

\runninghead{Progress on the Kretschmann-Schlingemann-Werner Conjecture}
            {Frederik vom Ende}

\normalsize\textlineskip
\thispagestyle{empty}
\setcounter{page}{1}

\copyrightheading{0}{0}{2003}{000--000}

\vspace*{0.88truein}

\alphfootnote

\fpage{1}

\centerline{\bf
PROGRESS ON THE KRETSCHMANN-SCHLINGEMANN-WERNER}
\vspace*{0.035truein}
\centerline{\bf CONJECTURE
}
\vspace*{0.37truein}
\centerline{\footnotesize
FREDERIK VOM ENDE
}
\vspace*{0.015truein}
\centerline{\footnotesize\it Dahlem Center for Complex Quantum Systems, Freie Universit\"at Berlin, Arnimallee 14}
\baselineskip=10pt
\centerline{\footnotesize\it Berlin, 14195,
Germany
}
\vspace*{0.225truein}
\publisher{(received date)}{(revised date)}

\vspace*{0.21truein}

\abstracts{
Given any pair of quantum channels $\Phi_1,\Phi_2$ such that at least one of them has Kraus rank one, as well as any respective Stinespring isometries
$V_1,V_2$,
we prove that there exists a unitary $U$ on the environment such that
$\|V_1-(\mathbbm1\otimes U)V_2\|_\infty\leq\sqrt{2\|\Phi_1-\Phi_2\|_\diamond}$.
Moreover, we provide a simple example which shows that the factor $\sqrt2$ on the right-hand side is optimal,
and we conjecture that this inequality holds for every pair of channels.
}{}{}

\vspace*{10pt}

\keywords{Stinespring dilation, Stinespring isometries, quantum channels, diamond norm}
\vspace*{3pt}
\communicate{to be filled by the Editorial}

\vspace*{1pt}\textlineskip    
\section{Introduction}
A well-known consequence of Stinespring's dilation theorem \cite{Stinespring55}
is that every quantum channel arises from some action on a larger system. More precisely, for every completely positive trace-preserving map 
there exists a Hilbert space (representing the environment)
as well as an isometry $V$---mapping the input space of the channel to the output space coupled to the environment---such that the original channel
is recovered by tracing the environment out of $V(\cdot)V^*$ \cite[Thm.~6.9]{Holevo12}.
Equivalently, every quantum channel can be expressed in operator-sum form
using so-called Kraus operators \cite{Kraus71}.
Both these representations of quantum channels are ubiquitous in---and fundamental to---quantum information and quantum computation \cite{Watrous18}.

While every such $V$ (called \textit{Stinespring isometry}) induces a unique quantum channel via ${\rm tr}_E(V(\cdot)V^*)$, every channel admits uncountably many Stinespring isometries, even after restricting the dimension of the environment Hilbert space.
This is why Kretschmann et al.~\cite{Kretschmann08} posed
the question whether any two channels that are, in some sense, ``close together'' admit Stinespring isometries which are also ``close together''.
What they were able to show is that for any two quantum channels $\Phi_1,\Phi_2:\mathbb C^{n\times n}\to\mathbb C^{k\times k}$ there \textit{exist} Stinespring isometries $V_1,V_2$ with common dilation space such that
\begin{equation}\label{eq:intro1}
\|V_1-V_2\|_\infty\leq\sqrt{\|\Phi_1-\Phi_2\|_\diamond}\,,
\end{equation}
under the crucial assumption that the dimension of the dilation space is at least $2nk$.
For the precise statement refer to \cite[Thm.~1]{Kretschmann08} or \cite[Prop.~5]{DAriano07}; a slight refinement which lowers the necessary environment dimension is given in Proposition~1 
below.

This result has since been applied to various fields of quantum physics:
Already in their original paper Kretschmann et al.~used it to derive an information-disturbance tradeoff for quantum channels which has since been employed in quantum computation \cite{LRR19}, (approximate) quantum error correction \cite{BO10,CKLT16}, and, subsequently, holography \cite{KK17,HP19,AP22,FL22}.
Moreover, the above continuity result has been used to answer questions, e.g., in quantum communication \cite{HW12,LKDW18}, quantum simulation \cite{BO11},
error correction \cite{KKS08}, and cryptography \cite{DAriano07,AGM21}.
Finally, this result has since been generalized to $C^*$-algebras \cite{Kretschmann08_2} as well as to energy-constrained channels \cite{Shirokov19,Shirokov20,Shirokov22}, which in turn has been used to study convergence of quantum channels in the strong${}^*$ operator topology \cite{Shirokov21}.

In this work we will focus on the assumption of Kretschmann et al.'s result that the dilation space has to be ``large enough'';
indeed, they follow up on their result by saying that they ``do not claim that for \textit{any} common dilation space
there exist isometries such that [Eq.~\eqref{eq:intro1} holds but they]
conjecture this to be true'' \cite[p.~1711]{Kretschmann08}.
While we will provide a counterexample down below (cf.~Example~1) 
there is much
structure to uncover here.
First---based on said example---we formulate the (updated) Kretschmann-Schlingemann-Werner conjecture:\medskip

{\noindent\bf Conjecture~1} {\it Let completely positive, trace-preserving linear maps $\Phi_1,\Phi_2:\mathbb C^{n\times n}\to\mathbb C^{k\times k}$, $n,k\in\mathbb N$ be given.
Then for all $m\in\mathbb N$ and for all Stinespring isometries $V_1,V_2:\mathbb C^n\to\mathbb C^k\otimes\mathbb C^m$ of $\Phi_1,\Phi_2$, respectively,
it holds that }
\begin{equation}\label{eq:conj_KSW08_update}
\min_{U\in\mathsf U(m)}\|V_1-(\mathbbm1\otimes U)V_2\|_\infty\leq \sqrt{2\|\Phi_1-\Phi_2\|_\diamond}\,.
\end{equation}

\noindent 
At this point we note that the ``converse'' inequality
$$
\|\Phi_1-\Phi_2\|_\diamond\leq 2\min_{U\in\mathsf U(m)}\|V_1-(\mathbbm1\otimes U)V_2\|_\infty\,,
$$
was obtained in
\cite{Kretschmann08}\footnote{
To recap their argument, given any $m,n,k\in\mathbb N$ and arbitrary linear maps $V_1,V_2:\mathbb C^n\to\mathbb C^k\otimes\mathbb C^m$ one has ${\rm id}\otimes({\rm tr}_{\mathbb C^m}(V_1(\cdot)V_1^*-V_2(\cdot)V_2^*))={\rm tr}_{\mathbb C^m}((\mathbbm1\otimes V_1)(\cdot)(\mathbbm1\otimes V_1)^*-(\mathbbm1\otimes V_2)(\cdot)(\mathbbm1\otimes V_2)^*)$. Then the operator norm of the latter---which equals the 
diamond norm of ${\rm tr}_{\mathbb C^m}(V_1(\cdot)V_1^*)-{\rm tr}_{\mathbb C^m}(V_2(\cdot)V_2^*)$---can be upper bounded by $(\|V_1\|_\infty+\|V_2\|_\infty)\|V_1-V_2\|_\infty$
via the triangle inequality. For isometries this equals $2\|V_1-V_2\|_\infty$, 
and this continues to be an upper bound when taking the minimum over the 
local unitary orbits.
}
${}$
without restriction on $m$, so proving Conjecture~1 
would establish full equivalence of the diamond norm and this ``distance'' (cf.~also the remark at the end of Sec.~\ref{sec:main})
between Stinespring isometries.
Aside from showing that the factor $\sqrt2$ on the right-hand side of~\eqref{eq:conj_KSW08_update} cannot be chosen any lower,
this paper's main contribution reads as follows.\medskip

{\noindent\bf Theorem~1 (Informal)}
{\it If $\Phi_1$ or $\Phi_2$ has Kraus rank one, then Conjecture~1 
holds.
This includes the case where at least one of the two channels is unitary.}
\medskip

The idea of our proof boils down to treating the following two cases:
If the two isometries $V_1,V_2$ are ``similar enough'' (i.e.~close in norm up to  local unitaries), then the minimum in Eq.~\eqref{eq:conj_KSW08_update} has an explicit form via the operational fidelity of $\Phi_1,\Phi_2$; note that this only holds in general because one of the channels has Kraus rank one.
Either way, the analytic expression of the minimum lets us show Eq.~\eqref{eq:conj_KSW08_update}
via an established inequality
that upper bounds the operational fidelity using the diamond norm.
On the other hand, if the (local unitary orbits of the) two isometries $V_1,V_2$ are ``far apart'', then the diamond norm distance between $\Phi_1$ and $\Phi_2$ is maximal meaning Eq.~\eqref{eq:conj_KSW08_update} is satisfied trivially. This implication, again, only holds because of the Kraus-rank one condition.


Now while Conjecture~1 
is supported by numerics, any proof of it has to differ from our proof (of Thm.~1) 
as the latter does not generalize to arbitrary channels.
We will address this in the outlook section where we also present some ideas on how one could tackle Conjecture~1 
beyond the assumption from Thm.~1. 

\section{Main Results}\label{sec:main}

As mentioned before, Kretschmann et al.~eventually extended their continuity result to $C^*$-algebras \cite{Kretschmann08_2}; yet, in this article we will work exclusively in finite dimensions.
First some notation: 
The collection of all linear maps $:\mathbb C^{n\times n}\to\mathbb C^{k\times k}$ will be denoted by $\mathcal L(\mathbb C^{n\times n},\mathbb C^{k\times k})$, while $\mathsf{CPTP}(n,k)$ refers to the subset of completely positive, trace-preserving maps (also called \textit{quantum channels} or \textit{quantum maps}).
Each channel $\Phi\in\mathsf{CPTP}(n,k)$ admits a \textit{Kraus rank} (sometimes also called \textit{Choi rank}) which is defined as the smallest number $r\in\mathbb N$ for which there exist $\{K_j\}_{j=1}^r\subset\mathbb C^{k\times n}$ such that $\Phi\equiv\sum_{j=1}^rK_j(\cdot)K_j^*$. In particular it holds that $r\leq nk$, cf.~\cite[Ch.~2.2.2]{Watrous18}.

One channel that will be particularly important is the partial trace (in the Schr\"odinger picture) ${\rm tr}_{\mathbb C^m}:\mathbb C^{k\times k}\otimes\mathbb C^{m\times m}\to\mathbb C^{k\times k}$, that is, ${\rm tr}_{\mathbb C^m}$ is the unique linear map which satisfies ${\rm tr}({\rm tr}_{\mathbb C^m}(A)B)={\rm tr}(A(B\otimes\mathbbm1_m))$ for all $A\in\mathbb C^{k\times k}\otimes\mathbb C^{m\times m}$, $B\in\mathbb C^{k\times k}$.
Next, we denote by
$\mathbb D(\mathbb C^n)$ the
set of all $n$-level quantum states (i.e.~positive semi-definite matrices of unit trace)
which, as usual, is equipped with the trace norm $\|\cdot\|_1$ (i.e.~the sum of all singular values of the input).
In contrast, $\|\cdot\|_\infty$ refers to the operator norm (on matrices) which is given by the largest singular value of the input.
Finally, $\mathsf U(n)$ is the Lie group of all unitary $n\times n$ matrices and
$\mathfrak u(n)$ is its Lie algebra, that is, $i\mathfrak u(n)$ is the collection of all Hermitian $n\times n$ matrices.

With this we come to the measures employed in this paper to ``compare'' channels.
The main tool certainly is the \textit{diamond norm} $\|\cdot\|_\diamond$---also called completely bounded trace norm---which for any $\Phi\in\mathcal L(\mathbb C^{n\times n},\mathbb C^{k\times k})$ is defined as the supremum of $\|({\rm id}_n
\otimes\Phi)(A)\|_1$ taken over all $A\in\mathbb C^{n\times n}\otimes\mathbb C^{n\times n}$ with $\|A\|_1=1$, cf.~\cite[Def.~3.43 ff.]{Watrous18}.
Moreover, we will use the \textit{operational fidelity}
\begin{align*}
F(\Phi_1,\Phi_2):=&\min_{\psi\in\mathbb C^n\otimes\mathbb C^n,\|\psi\|= 1}f\big( ({\rm id}\otimes\Phi_1)(|\psi\rangle\langle\psi|),({\rm id}\otimes\Phi_2)(|\psi\rangle\langle\psi|) \big)\\
=&\min_{\rho\in\mathbb D(\mathbb C^n\otimes\mathbb C^n)}f\big( ({\rm id}\otimes\Phi_1)(\rho),({\rm id}\otimes\Phi_2)(\rho) \big)
\end{align*}
where $f(\rho,\sigma):={\rm tr}(\sqrt{\sqrt{\rho}\sigma\sqrt{\rho}})$ is the usual fidelity, cf.~\cite[Eq.~(52)]{Belavkin05}.
A key characterization of the operational fidelity
is that 
for any two isometries $V_1,V_2:\mathbb C^n\to\mathbb C^k\otimes\mathbb C^m$
one has
\begin{equation}\label{eq:fid_char}
\begin{split}
F({\rm tr}_{\mathbb C^m}(V_1(\cdot)V_1^*),{\rm tr}_{\mathbb C^m}(V_2(\cdot)V_2^*))&=\max_{\substack{W\in\mathbb C^{m\times m}\\\|W\|_\infty\leq 1}}\min_{\rho\in\mathbb D(\mathbb C^n)}{\rm Re}({\rm tr}(\rho V_1^*(\mathbbm1\otimes W)V_2))\\
&=\frac12\max_{\substack{W\in\mathbb C^{m\times m}\\\|W\|_\infty\leq 1}}\min\,{\rm spec}\,(V_1^*(\mathbbm1\otimes W)V_2+V_2^*(\mathbbm1\otimes W^*)V_1) \,, 
\end{split}
\end{equation}
where, here and henceforth, ${\rm spec}$ denotes the spectrum;
the first equality in~\eqref{eq:fid_char} is \cite[Eq.~(24)]{Kretschmann08}, and the second equality is readily verified.

Having set the stage we are now ready to recap (and slightly refine) the original continuity result of Kretschmann et al.
Not only is doing so the obvious starting point, but
the proof below and the techniques used therein---which
differ slightly from the original proof of Kretschmann et al., although it is in the same
spirit---will become important later
when we generalize the result.\medskip

{\noindent\bf Proposition~1} {\it Let $m,n,k\in\mathbb N$ and $\Phi_1,\Phi_2\in\mathsf{CPTP}(n,k)$ be given, and assume that $m$ is not smaller than the sum of the Kraus ranks of $\Phi_1$ and $\Phi_2$.
Then for all Stinespring isometries
$V_1,V_2:\mathbb C^n\to\mathbb C^k\otimes\mathbb C^m$ of $\Phi_1,\Phi_2$, respectively,
there exists $U\in\mathsf U(m)$ such that}
\begin{equation}\label{eq:coro_ksw08_0}
\|V_1-(\mathbbm1\otimes U)V_2\|_\infty^2=2(1-F(\Phi_1,\Phi_2))\,.
\end{equation}
{\it In particular, for any such $m$ and all such $V_1,V_2$ it holds that}
\begin{equation}\label{eq:coro_ksw08}
\min_{U\in\mathsf U(m)}\|V_1-(\mathbbm1\otimes U)V_2\|_\infty\leq \sqrt{\|\Phi_1-\Phi_2\|_\diamond}\,.
\end{equation}
\proof{
It suffices to prove~\eqref{eq:coro_ksw08_0} because then~\eqref{eq:coro_ksw08} follows from 
the Fuchs-van de Graaf inequality $2(1-F(\Phi_1,\Phi_2))\leq\|\Phi_1-\Phi_2\|_\diamond$ (cf.~\cite{FG99,Belavkin05} and \cite[Lemma~2]{Kretschmann08}).

The strategy for showing~\eqref{eq:coro_ksw08_0} is to first prove it for ``specific'' Stinespring isometries, and in a second step to relate it to the general case.
First, set $k_j:=\text{Kraus rank}\,(\Phi_j)$, $j=1,2$
and choose any Kraus operators $\{K_{1j}\}_{j=1}^{k_1}$, $\{K_{2j}\}_{j=1}^{k_2}\subset\mathbb C^{k\times n}$ of $\Phi_1,\Phi_2$, respectively.
With this define
$\tilde V_1,\tilde V_2:\mathbb C^n\to\mathbb C^k\otimes\mathbb C^m$
via $\tilde V_1x:=\sum_{j=1}^{k_1} K_{1j}x\otimes|j\rangle$, $\tilde V_2:=\sum_{j=1}^{k_2} K_{2j}x\otimes|k_1+j\rangle$ for all $x\in\mathbb C^n$;
because $m\geq k_1+k_2$ by assumption, $\tilde V_1,\tilde V_2$ are well-defined.
One readily verifies that $\tilde V_1,\tilde V_2$ are Stinespring isometries of $\Phi_1,\Phi_2$, respectively, and that for all $U\in\mathbb C^{m\times m}$
\begin{equation}\label{eq:V1V2_Kraus}
\tilde V_1^*(\mathbbm1\otimes U)\tilde V_2=\sum_{j=1}^{k_1}\sum_{\ell=1}^{k_2}K_{1j}^*K_{2\ell}\langle j|U|k_1+\ell\rangle\,.
\end{equation}
The reason~\eqref{eq:V1V2_Kraus} is important is
the fact that
\begin{align}
\|\tilde V_1-(\mathbbm1\otimes U)\tilde V_2\|_\infty^2&= \big\| \big(\tilde V_1-(\mathbbm1\otimes U)\tilde V_2\big)^*\big(\tilde V_1-(\mathbbm1\otimes U)\tilde V_2\big) \big\|_\infty \notag\\
&=\big\|2\cdot\mathbbm1-\tilde V_1^*(\mathbbm1\otimes U)\tilde V_2-\tilde V_2^*(\mathbbm1\otimes U^*)\tilde V_1\big\|_\infty\notag\\
&= 2-\min\,{\rm spec}\,\big(\tilde V_1^*(\mathbbm1\otimes U)\tilde V_2-\tilde V_2^*(\mathbbm1\otimes U^*)\tilde V_1\big)\label{eq:prop1_1}
\end{align}
for all $U\in\mathsf U(m)$,
where in the last step we used that $\tilde V_1^*(\mathbbm1\otimes U)\tilde V_2-\tilde V_2^*(\mathbbm1\otimes U^*)\tilde V_1$ is Hermitian (hence unitarily diagonalizable).
Now, writing any $U\in\mathbb C^{m\times m}$ in block form
$$
U=\begin{pmatrix}
U_{11}&U_{12}\\U_{21}&U_{22}
\end{pmatrix}
$$
with $U_{11}\in\mathbb C^{k_1\times k_1}$ (i.e.~$U_{12}\in\mathbb C^{k_1\times(m-k_1)}$)---together with
Eq.~\eqref{eq:V1V2_Kraus}---shows that only $U_{12}$ appears in $\tilde V_1^*(\mathbbm1\otimes U)\tilde V_2+\tilde V_2^*(\mathbbm1\otimes U^*)\tilde V_1$.
With this in mind let $W\in\mathbb C^{{m}\times {m}}$ be any matrix where
$
\max_{\|W\|_\infty\leq 1}\min\,{\rm spec}\,(\tilde V_1^*(\mathbbm1\otimes W)\tilde V_2+\tilde V_2^*(\mathbbm1\otimes W^*)\tilde V_1)
$
is attained.
Because $\|W\|_\infty\leq 1$, one in particular has $\|W_{12}\|_\infty\leq1$
so
$$
W_0:=\begin{pmatrix}
\sqrt{\mathbbm1-W_{12}W_{12}^*}&W_{12}\\
W_{12}^*&-\sqrt{\mathbbm1-W_{12}^*W_{12}}
\end{pmatrix}\in\mathbb C^{m\times m}
$$
is unitary, cf.~\cite[Appendix, Sec.~4]{SzNagy90}.
Thus $\min\,{\rm spec}\,(\tilde V_1^*(\mathbbm1\otimes W)\tilde V_2+\tilde V_2^*(\mathbbm1\otimes W^*)\tilde V_1)$ remains unchanged when replacing $W$ by $W_0$.
Using~\eqref{eq:prop1_1} this lets us conclude
\begin{align*}
\|\tilde V_1-(\mathbbm1\otimes W_0)\tilde V_2\|_\infty^2&\overset{\hphantom{\eqref{eq:fid_char}}}= 2-\min\,{\rm spec}\,\big(\tilde V_1^*(\mathbbm1\otimes W_0)\tilde V_2+\tilde V_2^*(\mathbbm1\otimes W_0^*)\tilde V_1\big)\\
&\overset{\hphantom{\eqref{eq:fid_char}}}= 2-\min\,{\rm spec}\,(\tilde V_1^*(\mathbbm1\otimes W)\tilde V_2+\tilde V_2^*(\mathbbm1\otimes W^*)\tilde V_1)\\
&\overset{\hphantom{\eqref{eq:fid_char}}}= 2-\max_{\|U\|_\infty\leq 1}\min\,{\rm spec}\,(\tilde V_1^*(\mathbbm1\otimes U)\tilde V_2+\tilde V_2^*(\mathbbm1\otimes U^*)\tilde V_1)\\
&\overset{\eqref{eq:fid_char}}=2(1-F(\Phi_1,\Phi_2))\,,
\end{align*}
that is, \eqref{eq:coro_ksw08_0} holds for $\tilde V_1$, $\tilde V_2$, and $U=W_0$.
Now for the general case: 
Given $m\in\mathbb N$, $m\geq k_1+k_2$ and any Stinespring isometries
$V_1,V_2:\mathbb C^n\to\mathbb C^k\otimes\mathbb C^m$ of $\Phi_1,\Phi_2$,
because $V_k$ and $\tilde V_k$ describe the same channel there exist $U_1,U_2\in\mathsf U(m)$ such that $V_j=(\mathbbm1\otimes U_j)\tilde V_j$, $j=1,2$ \cite[Coro.~2.24]{Watrous18}.
With this, setting
$U:=U_1^{*}W_0U_2$ yields
\begin{align*}
\|V_1-(\mathbbm1\otimes U)V_2\|_\infty&=\|(\mathbbm1\otimes U_1)V_1-(\mathbbm1\otimes W_0)(\mathbbm1\otimes U_2)V_2\|_\infty\\
&=\|\tilde V_1-(\mathbbm1\otimes W_0)\tilde V_2\|_\infty= \sqrt{2(1-F(\Phi_1,\Phi_2))}
\end{align*}
which concludes the proof.
}\medskip

\noindent We remark that the quantity $2(1-F(\Phi_1,\Phi_2))$ is also known as the \textit{Bures distance} of $\Phi_1$ and $\Phi_2$ \cite{Shirokov19}, in analogy to the Bures distance of quantum states, cf., e.g., \cite[Ch.~10.2.3]{Holevo12}.

It turns out---as already hinted at in the introduction---that the assumption on the dimension $m$ of the auxiliary space being ``large enough'' is necessary for~\eqref{eq:coro_ksw08}
(and thus for~\eqref{eq:coro_ksw08_0})
to hold.
Let us give an example to substantiate this:\medskip

{\noindent\bf Example~1} {\it Define the unitaries $V_1:=\mathbbm1$ and $V_2:={\rm diag}(e^{2\pi ij/n})_{j=1}^n$, $n\in\mathbb N$.
We claim that 
\begin{equation}\label{eq:prop_ksw_const_1}
\min_{z\in\mathsf U(1)}\|V_1-zV_2\|_\infty=\sin\Big(\frac{\pi (n-1)}{2n}\Big)\sqrt{2\|V_1(\cdot)V_1^*-V_2(\cdot)V_2^*\|_\diamond}\,.
\end{equation}
The case $n=1$ is trivial so w.l.o.g.~$n\geq 2$.
For the left-hand side of~\eqref{eq:prop_ksw_const_1}---using that the spectrum of $V_2$ is symmetric w.r.t.~the real axis---a simple geometric argument shows
\begin{align*}
\max_{\phi\in\mathbb R}\|\mathbbm1- e^{i\phi}V_2\|_\infty&=
\|\mathbbm1-V_2\|_\infty=\big|1-e^{i\pi(n-1)/n}\big|\\
&= \sqrt{2\Big( 1-\cos\Big( \frac{\pi(n-1)}{n} \Big) \Big)} =2\sin\Big(\frac{\pi (n-1)}{2n}\Big)
\end{align*}
where in the last step we used the standard half-angle formula.
On the other hand, \cite[Thm.~12]{JKP09} shows
$\|V_1(\cdot)V_1^*-V_2(\cdot)V_2^*\|_\diamond=2$
for our choice of $V_1,V_2$,
which altogether establishes~\eqref{eq:prop_ksw_const_1}.}
\medskip

\noindent Not only does this example show that Eq.~\eqref{eq:coro_ksw08}
cannot hold for all $m\in\mathbb N$
as soon as $n\geq 3$, it even
shows that no ``universal'' (i.e.~dimension-independent) constant in Conjecture~1 
can 
be smaller than $\sqrt2$ (as $\sin(\frac{\pi (n-1)}{2n})\to 1$ for $n\to\infty$).
In fact this ``bound'' of $\sqrt2$ is attained in infinite dimensions: set
$\mathcal H:=\ell^2(\mathbb Z)$, $V_1:=\mathbbm1$, and
$V_2:=\sum_{n\in\mathbb Z}|n\rangle\langle n+1|\in\mathsf U(\mathcal H)$ is the bilateral shift.
Moreover, numerics suggest that Example~1 
describes the maximal violation of Eq.~\eqref{eq:coro_ksw08} in each dimension.

Altogether, this example is what motivates Conjecture~1, 
and at this point we are ready to prove the latter for the special case where at least one of the channels has Kraus rank one.
To do so we first need the following lemma;
it appears similar to Eq.~\eqref{eq:coro_ksw08_0}
but it does not feature any assumption on $m$, hence why
the proof below differs substantially from the proof of Proposition~1. 
\medskip

\begin{lemma}\label{lemma0}
Let $m,n,k\in\mathbb N$, $U\in\mathbb C^{k\times n}$, a vector $\phi\in\mathbb C^m$ with $\|\phi\|=1$, and a linear map $V:\mathbb C^n\to\mathbb C^k\otimes\mathbb C^m$ be given such that $U$ and $V$ are isometries.
If there exists $W\in\mathsf U(m)$ such that
$\|U\otimes|\phi \rangle-(\mathbbm1\otimes W)V\|_\infty\leq\sqrt 2$, then
\begin{equation}\label{eq:thm0_1}
\min_{W\in\mathsf U(m)}\|U\otimes|\phi \rangle-(\mathbbm1\otimes W)V\|_\infty^2={2\big(1-F\big(U(\cdot)U^*,{\rm tr}_{\mathbb C^m}(V(\cdot)V^*)\big)\big)}\,.
\end{equation}
\end{lemma}
\proof{
The key is the (non-linear) functional
\begin{equation}\label{eq:nonlinfunct}
\begin{split}
\Gamma:\mathbb C^{m}&\to\mathbb R\\
\psi&\mapsto\min\,{\rm spec}\,\big((U^*\otimes\langle \psi|)V+V^*(U\otimes |\psi\rangle) \big)
\end{split}
\end{equation}
and the observation that $\Gamma(\lambda \psi)=\lambda \Gamma(\psi)$ for all $\psi\in\mathbb C^{m}$ and all $\lambda\geq 0$.
Now by assumption there exists $W\in\mathsf U(m)$ such that
$\|U\otimes|\phi \rangle-(\mathbbm1\otimes W)V\|_\infty\leq\sqrt 2$; by~\eqref{eq:prop1_1} this implies that
$(U^*\otimes\langle \phi |W)V+V^*(U\otimes W^*|\phi \rangle)$
is positive semi-definite.
In particular this shows that
\begin{equation}\label{eq:max_unitsphere_nonneg}
\max_{\|\psi\|=1}\Gamma(\psi)\geq\Gamma(W^*|\phi\rangle)\geq 0
\end{equation}
meaning
the maximum of $\Gamma(\psi)$ over the unit ball is equal to the maximum over the unit sphere:
To see this, let $\psi_{\rm max}\in\mathbb C^n$, $\|\psi_{\rm max}\|\leq 1$ be any vector such that $\Gamma(\psi_{\rm max})=\max_{\|\psi\|\leq1}\Gamma(\psi)$.
Then\footnote{
We may assume w.l.o.g.~that $\psi_{\rm max}\neq 0$:
If the maximum of $\Gamma$ over the unit ball were zero, then
by~\eqref{eq:max_unitsphere_nonneg} $\max_{\|\psi\|\leq1}\Gamma(\psi)=0\leq \Gamma(W^*|\phi\rangle)\leq 
\max_{\|\psi\|=1}\Gamma(\psi)\leq 
\max_{\|\psi\|\leq1}\Gamma(\psi)$ as desired.
}
\begin{equation}\label{eq:lemma0_1}
\max_{\|\psi\|\leq1}\Gamma(\psi)=\Gamma(\psi_{\rm max})=\|\psi_{\rm max}\|\Gamma\Big(\frac{\psi_{\rm max}}{\|\psi_{\rm max}\|}\Big)\leq \Gamma\Big(\frac{\psi_{\rm max}}{\|\psi_{\rm max}\|}\Big)\leq 
\max_{\|\psi\|=1}\Gamma(\psi)\leq 
\max_{\|\psi\|\leq1}\Gamma(\psi)\,,
\end{equation}
where in the in the third step we used that $\Gamma(\psi_{\rm max})\geq 0$.
This lets us calculate
\begin{align*}
2\big(1-F\big(U(\cdot)U^*,{\rm tr}_{\mathbb C^m}(V(\cdot)V^*)\big)\big)
&\overset{\eqref{eq:fid_char}}=2-\max_{\|W\|_\infty\leq 1}\min\,{\rm spec}\,(U^*\otimes\langle \phi |W)V+V^*(U\otimes W^*|\phi \rangle) \\
&\overset{\hphantom{\eqref{eq:prop1_1}}}=2-\max_{\|\psi\|\leq 1}\Gamma(\psi)\overset{\eqref{eq:lemma0_1}}=2-\max_{\|\psi\|= 1}\Gamma(\psi)
\\
&\overset{\hphantom{\eqref{eq:prop1_1}}}=\min_{W\in\mathsf U(m)}\big(2-\Gamma(W^*|\phi \rangle)\big)\\
&\overset{\eqref{eq:prop1_1}}=\min_{W\in\mathsf U(m)}\|U\otimes|\phi \rangle-(\mathbbm1\otimes W)V\|_\infty^2
\end{align*}
which concludes the proof.
}\medskip

The condition from Lemma~\ref{lemma0} that the (local unitary orbits of the) isometries are no more than $\sqrt2$ apart already appeared, implicitly,
in the proof of Proposition~1: 
there, Eq.~\eqref{eq:V1V2_Kraus} implies $\tilde V_1^*\tilde V_2=0$ and thus $\|\tilde V_1-\tilde V_2\|_\infty=\sqrt 2$.
However, we emphasize that 
there do exist isometries (the orbits of) which are less than $\sqrt2$ apart but
the square of said distance is \textit{not} equal to---but rather \textit{strictly} 
larger than---$2(1-F({\rm tr}_{\mathbb C^m}(V_1(\cdot)V_1^*),{\rm tr}_{\mathbb C^m}(V_2(\cdot)V_2^*)))$;
an example is given in Appendix~A.
In particular this means that Lemma~\ref{lemma0} fails for general isometries.

Either way, with Lemma~\ref{lemma0} at hand we are ready to prove this paper's main result.\medskip

{\noindent\bf Theorem~1.} 
{\it Let $m,n,k\in\mathbb N$ and $U\in\mathbb C^{k\times n}$ with $U^*U=\mathbbm1$ be given.
If $V_1:\mathbb C^n\to\mathbb C^k\otimes\mathbb C^m$ is any Stinespring isometry of $U(\cdot)U^*$, and if $V_2:\mathbb C^n\to\mathbb C^k\otimes\mathbb C^m$ is an arbitrary isometry,
then
\begin{equation}\label{eq:thm0_formal}
\min_{W\in\mathsf U(m)}\|V_1-(\mathbbm1\otimes W)V_2\|_\infty\leq \sqrt{2\|U(\cdot)U^*-{\rm tr}_{\mathbb C^m}(V_2(\cdot)V_2^*)\|_\diamond}\,.
\end{equation}}
\proof{
Note that given any $\phi\in\mathbb C^m$ with $\|\phi\|=1$, $U\otimes|\phi\rangle$ is a Stinespring isometry of $U(\cdot)U^*$ and, moreover, by \cite[Coro.~2.24]{Watrous18} every Stinespring isometry of $U(\cdot)U^*$ is of this form.
For what follows we adapt the notation from the proof of Lemma~\ref{lemma0}.
We distinguish two cases:
\begin{itemize}
\item[1.] If $\min_{W\in\mathsf U(m)}\|U\otimes|\phi\rangle-(\mathbbm1\otimes W)V_2\|_\infty\leq\sqrt2$, then Lemma~\ref{lemma0} together with Fuchs-van de Graaf (i.e.~$2(1-F(\Phi_1,\Phi_2))\leq\|\Phi_1-\Phi_2\|_\diamond$, cf.~proof of Proposition~1) 
readily
implies~\eqref{eq:thm0_formal}.
\item[2.] 
Assume now that $\min_{W\in\mathsf U(m)}\|U\otimes|\phi\rangle-(\mathbbm1\otimes W)V_2\|_\infty>\sqrt2$.
By Eq.~\eqref{eq:prop1_1} this implies that
$(U^*\otimes\langle \phi |W)V+V^*(U\otimes W^*|\phi \rangle)$ has a negative eigenvalue for all $W\in\mathsf U(m)$.
Thus the functional $\Gamma$ from Eq.~\eqref{eq:nonlinfunct} takes only negative values on the unit sphere;
but $\Gamma$ also satisfies $\Gamma(\lambda(\cdot))\equiv\lambda \Gamma$ for all $\lambda\geq 0$, so $\Gamma$ is negative on all of $\mathbb C^m\setminus\{0\}$. This shows
$$
F\big(U(\cdot)U^*,{\rm tr}_{\mathbb C^m}(V(\cdot)V^*)\big)\overset{\eqref{eq:fid_char}}=\frac12\max_{\|\psi\|\leq 1}\Gamma(\psi)=\Gamma(0)=0\,.
$$
Thus, by Fuchs-van de Graaf \cite[Lemma~2]{Kretschmann08} $\|U(\cdot)U^*-{\rm tr}_{\mathbb C^m}(V(\cdot)V^*)\|_\diamond=2$. Altogether this implies the claim as the l.h.s.~of~\eqref{eq:thm0_formal} is trivially upper bounded by $2$ as a consequence of the triangle inequality.
\end{itemize}
}\medskip

Again we point out that this proof strategy cannot cover all of Conjecture~1: 
Not only does it rely on (some version of) Eq.~\eqref{eq:coro_ksw08_0}---which is not valid in general, cf.~Appendix~A---but
$\min_{W\in\mathsf U(m)}\|V_1-(\mathbbm1\otimes W)V_2\|_\infty>\sqrt2$
does in general \textit{not} imply that the channels ${\rm tr}_{\mathbb C^m}(V_j(\cdot)V_j^*)$, $j=1,2$ have zero fidelity, refer to Appendix~B for an example.
However, this does not rule out the possibility of modifying this paper's ideas and techniques to fit the case of general isometries; more on this in Sec.~\ref{sec_con_out}.

Finally, a word of caution: The way we measure how far two Stinespring isometries ``are apart'' in Conjecture~1 
is \textit{not} a metric on the set of quantum channels, at least not without further ado.
Given any $\Phi_1,\Phi_2\in\mathsf{CPTP}(n,k)$ set $m:=\max_{j=1,2}\text{Kraus rank}\,(\Phi_j)$. Then, given any Stinespring isometries $V_1,V_2:\mathbb C^n\to\mathbb C^k\otimes\mathbb C^m$ of $\Phi_1,\Phi_2$
define
\begin{equation}\label{eq:def_d_stine}
d(\Phi_1,\Phi_2):=\min_{U\in\mathsf U(m)}\|V_1-(\mathbbm1\otimes U)V_2\|_\infty\,.
\end{equation}
By \cite[Coro.~2.24]{Watrous18} $d$ does not depend on the particular choice of $V_1,V_2$, that is, the map $d:\mathsf{CPTP}(n,k)\times\mathsf{CPTP}(n,k)\to[0,\infty)$ 
is well-defined.
While $d$ is symmetric and positive definite, it does not satisfy the triangle inequality.
A counter-example is given by $n=k=3$,
$\Phi_1={\rm id}$, $\Phi_2=U(\cdot)U^*$ with $U:={\rm diag}(1,e^{2\pi i/3},e^{4\pi i/3})$, as well as $\Phi_3:=\frac12(\Phi_1+\Phi_2)$:
then
$d(\Phi_1,\Phi_2)=\sqrt3$ (by Example~1) 
while 
\begin{align*}
d(\Phi_1,\Phi_3)&\leq \big\|\mathbbm1\otimes |0\rangle-\tfrac{\mathbbm1\otimes |0\rangle+U\otimes|1\rangle}{\sqrt2}\big\|_\infty=\sqrt{2-\sqrt2}\\
d(\Phi_2,\Phi_3)&\leq \big\|U\otimes |0\rangle-(\mathbbm1\otimes\sigma_x)\tfrac{\mathbbm1\otimes |0\rangle+U\otimes|1\rangle}{\sqrt2}\big\|_\infty= \big\|U\otimes |0\rangle-\tfrac{\mathbbm1\otimes |1\rangle+U\otimes|0\rangle}{\sqrt2}\big\|_\infty=\sqrt{2-\sqrt2}\,.
\end{align*}
This leads to the desired contradiction $$d(\Phi_1,\Phi_2)=\sqrt3>2\sqrt{2-\sqrt2}\geq d(\Phi_1,\Phi_3)+d(\Phi_3,\Phi_2)\,.$$
Note that this example relies on $\Phi_3$ having a different Kraus rank than $\Phi_1,\Phi_2$.
Indeed,
one possibility to resolve this issue could be to include the number $m$ in the minimum in~\eqref{eq:def_d_stine}.
However, doing so might make such a quantity unfeasible for studying general channels
as the effect of the auxiliary dimension $m$ on $\min_{U\in\mathsf U(m)}\|V_1-(\mathbbm1\otimes U)V_2\|_\infty$ is not well understood.

\section{Conclusions and Outlook}\label{sec_con_out}

We updated the Kretschmann-Schlingemann-Werner conjecture (stated as Conjecture~1) 
in the introduction) based on a counterexample (Example~1), 
and we went on to prove the special case where at least one of the channels has Kraus rank one; this includes the case where at least one of the channels is unitary.
The main idea of our proof was to show that, in this case, $\min_{U\in\mathsf U(m)}\|V_1-(\mathbbm1\otimes U)V_2\|_\infty^2$
can be characterized via the Bures distance (resp.~the operational fidelity) of the induced channels, assuming this minimum does not exceed $2$.
Generally, this minimum can be strictly larger than $2(1-F(\Phi_1,\Phi_2))$ (cf.~Appendix~A) which, however, does not disprove the conjecture,
but rather shows that our proof strategy has to be modified.
One idea---assuming one wants to keep employing the Fuchs-van de Graaf inequalities---could be to show
\begin{equation}\label{eq:outlook_1}
2\max_{\|U\|_\infty\leq 1}\Gamma(U)\leq 2+\max_{U\in\mathsf U(m)}\Gamma(U)
\end{equation}
for all $m,n,k\in\mathbb N$ and all isometries $V_1,V_2\in\mathbb C^n\to\mathbb C^k\otimes\mathbb C^m$, where
$\Gamma:\mathbb C^{m\times m}\to\mathbb R$ is the non-linear functional $\Gamma(X):=\min\,{\rm spec}\,(V_1^*(\mathbbm1\otimes W)V_2+V_2^*(\mathbbm1\otimes W^*)V_1)$.
By Eqs.~\eqref{eq:fid_char} \& \eqref{eq:prop1_1} proving~\eqref{eq:outlook_1} would be equivalent to proving
$\min_{U\in\mathsf U(m)}\|V_1-(\mathbbm1\otimes U)V_2\|_\infty\leq 2\sqrt{1-F(\Phi_1,\Phi_2)}$.

Alternatively
one could study the effect the environment dimension $m$ has on the quantity $\min_{U\in\mathsf U(m)}\|V_1-(\mathbbm1\otimes U)V_2\|_\infty\,$;
after all---by the original result of Kretschmann et al.---Conjecture~1 
(even a stricter version thereof) holds as soon as $m$ is ``large enough'', cf.~also Proposition~1. 
More precisely, given $m,n,k\in\mathbb N$ and isometries $V_1,V_2:\mathbb C^n\to\mathbb C^k\otimes\mathbb C^m$, if one could show that
for all $m'> m$
\begin{equation}\label{eq:open_1}
\min_{U\in\mathsf U(m)}\|V_1-(\mathbbm1\otimes U)V_2\|_\infty
\leq\sqrt2\min_{U'\in\mathsf U(m')}\|(\mathbbm1\otimes\iota_{m'})V_1-(\mathbbm1\otimes U')(\mathbbm1\otimes \iota_{m'})V_2\|_\infty
\end{equation}
where the isometry $\iota_{m'}:\mathbb C^m\to\mathbb C^{m'}$ is defined via $\iota_{m'}(x):=(x,0_{m'-m})^\top$, then combining
Proposition~1 
with Eq.~\eqref{eq:open_1}
would imply Conjecture~1. 

While both ideas seem promising and could be directions for future research, for now the Kretschmann-Schlingemann-Werner conjecture remains open.
Let us conclude by presenting two related open questions: 
\begin{itemize}
\item As the bound in Conjecture~1 
is independent of the dimension of the underlying spaces one may even extend the former to infinite dimensions:
Given complex Hilbert spaces $\mathcal H,\mathcal K,\mathcal Z$ and arbitrary isometries $V_1,V_2:\mathcal H\to\mathcal K\otimes\mathcal Z$ does it hold that
\begin{equation*}
\min_{U\in\mathsf U(\mathcal Z)}\|V_1-(\mathbbm1_{\mathcal K}\otimes U)V_2\|_\infty\leq \sqrt{2\|{\rm tr}_{\mathcal Z}(V_1(\cdot)V_1^*)-{\rm tr}_{\mathcal Z}(V_2(\cdot)V_2^*)\|_\diamond}\,?
\end{equation*}
\item Given dynamic processes $\mathsf\Phi_1,\mathsf\Phi_2$ do there exist $m\in\mathbb N$ as well as sufficiently regular---but at least locally absolutely continuous---curves of Stinespring isometries\footnote{
In this context this means that $\mathsf V_1(t),\mathsf V_2(t):\mathbb C^n\to\mathbb C^k\otimes\mathbb C^m$ is a Stinespring isometry of $\mathsf\Phi_1(t),\mathsf\Phi_2(t)$ for all times $t$.
}
${}$
$\mathsf V_1,\mathsf V_2$ such that
\begin{equation}\label{eq:open_2}
\|\mathsf V_1(t)-\mathsf V_2(t)\|_{\infty}\leq\sqrt{2\|\mathsf\Phi_1(t)-\mathsf\Phi_2(t)\|_{\diamond}}
\end{equation}
for all $t$?
Be aware that~\eqref{eq:open_2} would \textit{not} be a direct consequence of Conjecture~1 
due to the additional continuity requirement on $\mathsf V_1,\mathsf V_2$.
This question is inspired by recent work of ours 
\cite{vomEnde23_StinespringApprox}
which explored the concept of ``dynamic'' Stinespring representations
and established at least the approximate existence of such \cite[Thm.~1]{vomEnde23_StinespringApprox}.
\end{itemize}

\nonumsection{Acknowledgements}
\noindent
I am grateful to Sumeet Khatri for the valuable discussions during the preparation of this manuscript
as well as the anonymous referee for their comments.
Moreover, I would like to thank Lennart Bittel for noticing that the proof of Proposition~1 
works not only when $m$ is not smaller than than twice the larger Kraus rank, but already if $m$ is the sum of the Kraus ranks.
This work has been supported by the Einstein Foundation (Einstein Research Unit on Quantum Devices) and the MATH+ Cluster of Excellence.

\nonumsection{References}

\appendix{: Counterexample to a Generalization of Lemma~\ref{lemma0}}

\noindent
Consider the isometries
$$
V_1:=\frac{1}{\sqrt3}\begin{pmatrix}
1&0\\0&1\\1&1\\1&-1
\end{pmatrix}\quad\text{ and }\quad
V_2:=\frac{1}{\sqrt3}\begin{pmatrix}
1&1\\1&-1\\1&0\\0&-1
\end{pmatrix}\,.
$$
The corresponding channels $\Phi_j:={\rm tr}_{\mathbb C^2}(V_j(\cdot)V_j^*)$, $j=1,2$ act on any $\rho\in\mathbb C^{2\times 2}$ like
\begin{align*}
\Phi_1(\rho)&= \begin{pmatrix}
 \frac{1}{3}(\rho_{11}+\rho_{22}) & \frac{1}{3} (\rho_{11}+\rho_{12}+\rho_{21}-\rho_{22}) \\
 \frac{1}{3} (\rho_{11}+\rho_{12}+\rho_{21}-\rho_{22}) & \frac{2}{3}  (\rho_{11}+\rho_{22})
\end{pmatrix} \\
\Phi_2(\rho)&=  \begin{pmatrix}
 \frac{2}{3}(\rho_{11}+\rho_{22}) & \frac{1}{3} (\rho_{11}-\rho_{12}+\rho_{21}+\rho_{22}) \\
 \frac{1}{3} (\rho_{11}+\rho_{12}-\rho_{21}+\rho_{22}) & \frac{1}{3}  (\rho_{11}+\rho_{22})
 \end{pmatrix}\,.
\end{align*}
We claim that
\begin{equation}\label{eq:appA_1}
\begin{split}
\max_{U\in\mathsf U(2)}\min\,{\rm spec}\,\big( &V_1^*(\mathbbm1\otimes U)V_2+V_2^*(\mathbbm1\otimes U^*)V_1 \big)=0.3805\ldots\\
&<\frac23\leq 
\max_{\|U\|_\infty\leq 1}\min\,{\rm spec}\,\big( V_1^*(\mathbbm1\otimes U)V_2+V_2^*(\mathbbm1\otimes U^*)V_1\big)
\end{split}
\end{equation}
which by the proof of Lemma~\ref{lemma0} is equivalent to 
\begin{align*}
\min_{U\in\mathsf U(2)}\|V_1-(\mathbbm1&\otimes U)V_2\|_\infty=\sqrt{2-0.3805\ldots}\\
&>\sqrt{2-\tfrac23}\geq\sqrt{2\big(1-F\big({\rm tr}_{\mathbb C^m}(V_1(\cdot)V_1^*),{\rm tr}_{\mathbb C^m}(V_2(\cdot)V_2^*)\big)\big)}\,.
\end{align*}
On the one hand $( V_1^*(\mathbbm1\otimes |2\rangle\langle 1|)V_2+V_2^*(\mathbbm1\otimes |1\rangle\langle 2|)V_1=\frac23\cdot\mathbbm1$; this gives the lower bound of the right-hand side of~\eqref{eq:appA_1}.
On the other hand consider an arbitrary element $U\in\mathsf U(2)$, that is,
$$
U=\begin{pmatrix}
 \cos (x)e^{i\zeta} &\sin (x)e^{i(\omega+\phi)} \\
-\sin (x)e^{-i\omega} &  \cos (x) e^{i(\phi-\zeta)}
\end{pmatrix}
$$
for some $x,\phi,\zeta,\omega\in\mathbb R$.
Then one finds:
\begin{align*}
&\min\,{\rm spec}\,\big( V_1^*(\mathbbm1\otimes U)V_2+V_2^*(\mathbbm1\otimes U^*)V_1\big)
=\frac23\Big( \cos (x) \cos (\zeta)-\cos (\omega) \sin (x)-\\
&\qquad{\sqrt{ 2+2 \cos ^2(x) \sin ^2(\phi-\zeta)-2 \cos ^2(x) \sin ^2(\zeta)-(\sin (x) \sin (\omega+\phi)-\cos (x) \sin (\phi-\zeta))^2}}\Big)
\end{align*}
Observing that setting $\zeta=\omega=0$ and $\phi=-\frac{\pi}{2}$ does not change the maximum, we may instead consider
$$
U=\begin{pmatrix}
 \cos (x) & i \sin (x) \\
 -\sin (x) & i \cos (x)
\end{pmatrix}
$$
with $x\in[0,\pi]$. Therefore
\begin{align*}
\max_{U\in\mathsf U(2)}\min\,{\rm spec}\,\big(&V_1^*(\mathbbm1\otimes U)V_2+V_2^*(\mathbbm1\otimes U^*)V_1 \big)\\&=\frac23\max_{x\in[0,\pi]}\big(\cos(x) - \sin(x) -\sqrt{2 + \cos(2x) + \sin(2 x)}\big)\\
&=\sqrt{\frac{16}9-2\sqrt{\frac{2}{3}}}\simeq 0.3805\ldots
\end{align*}
as claimed.

\appendix{: Example of Isometries That Are ``Far Apart'' but the Induced Channels Have Non-Zero Fidelity}
\noindent The following is a (numerical) example of two isometries $V_1,V_2:\mathbb C^3\to\mathbb C^3\otimes\mathbb C^2$ such that $\min_{U\in\mathsf U(m)}\|V_1-(\mathbbm1\otimes U)V_2\|_\infty>\sqrt2$, but
$F({\rm tr}_{\mathbb C^m}(V_1(\cdot)V_1^*),{\rm tr}_{\mathbb C^m}(V_2(\cdot)V_2^*))>0$ (equivalently by the Fuchs-van de Graaf inequalities: $\|{\rm tr}_{\mathbb C^m}(V_1(\cdot)V_1^*)-{\rm tr}_{\mathbb C^m}(V_2(\cdot)V_2^*)\|_\diamond<2$).
A random search yielded
\begin{align*}
V_1&=
\begin{pmatrix}
 0.0720257\, +0.403635 i & -0.27118-0.260568 i & -0.0507697+0.0669192 i \\
 0.19795\, +0.363156 i & 0.201747\, -0.0722566 i & 0.679242\, -0.0637345 i \\
 -0.259574+0.274006 i & -0.107846-0.210931 i & 0.0967138\, +0.297056 i \\
 -0.429354-0.410058 i & 0.0846627\, +0.124068 i & 0.559109\, -0.251464 i \\
 -0.335246+0.151762 i & -0.130565+0.325685 i & 0.111082\, +0.0316921 i \\
 -0.120023-0.12668 i & -0.446167-0.641698 i & 0.200846\, -0.0199049 i 
\end{pmatrix}\,,\\
V_2&=
\begin{pmatrix}
 -0.472877+0.283338 i & -0.0970526-0.387244 i & 0.0909608\, -0.338129 i \\
 -0.244363+0.193024 i & 0.00157303\, +0.23514 i & -0.104252-0.362985 i \\
 0.129531\, +0.238246 i & -0.0278328+0.419327 i & 0.00258036\, -0.341585 i \\
 0.245734\, +0.258541 i & -0.171596+0.148668 i & -0.0406064+0.0942153 i \\
 0.0945343\, +0.420365 i & 0.0985514\, +0.220795 i & 0.687703\, +0.354813 i \\
 0.461193\, -0.00493922 i & -0.38767-0.590272 i & 0.0784118\, +0.0507872 i
\end{pmatrix}\,;
\end{align*}
numerics claim that $\min_{U\in\mathsf U(2)}\|V_1-(\mathbbm1_3\otimes U)V_2\|_\infty=1.478>\sqrt2$
and that the minimum
is attained on the unitary
$$
U=\begin{pmatrix}
 -0.256631+0.0241997 i & -0.674035+0.692265 i \\
 0.156281\, +0.953484 i & -0.196551-0.166771 i 
\end{pmatrix}\,.
$$
Yet, the channels ${\rm tr}_{\mathbb C^m}(V_1(\cdot)V_1^*),{\rm tr}_{\mathbb C^m}(V_2(\cdot)V_2^*)$ have non-zero fidelity:
set $$
W_0:=\begin{pmatrix}
 -0.123603-0.0759052 i & -0.753418+0.567052 i \\
 -0.179898+0.274495 i & -0.137246+0.225905 i 
\end{pmatrix}
$$
and note $\|W_0\|_\infty=1$ as well as
\begin{align*}
F({\rm tr}_{\mathbb C^m}(V_1(\cdot)V_1^*),{\rm tr}_{\mathbb C^m}(V_2(\cdot)V_2^*))&\overset{\eqref{eq:fid_char}}=\frac12\max_{\substack{W\in\mathbb C^{m\times m}\\\|W\|_\infty\leq 1}}\min\,{\rm spec}\,(V_1^*(\mathbbm1\otimes W)V_2+V_2^*(\mathbbm1\otimes W^*)V_1)\\
&\overset{\hphantom{\eqref{eq:fid_char}}}\geq \frac12\min\,{\rm spec}\,(V_1^*(\mathbbm1\otimes W_0)V_2+V_2^*(\mathbbm1\otimes W_0^*)V_1)\simeq0.04\,.
\end{align*}

\end{document}